\documentclass[aps,prl,twocolumn,superscriptaddress,floatfix,10pt]{revtex4}
\usepackage{graphicx}
\usepackage{amsmath}
\usepackage{color}
\usepackage{wrapfig}
\begin{document}

\title{Reduced thermal conductivity in molecular forests}

\author{Aashish Bhardwaj}
\affiliation{Department of Mechanical Engineering, University of British Columbia, Vancouver BC V6T 1Z4, Canada}
\author{A. Srikantha Phani}
\affiliation{Department of Mechanical Engineering, University of British Columbia, Vancouver BC V6T 1Z4, Canada}
\author{Alireza Nojeh}
\affiliation{Department of Electrical and Computer Engineering,University of British Columbia, Vancouver BC V6T 1Z4, Canada}
\affiliation{Quantum Matter Institute, University of British Columbia, Vancouver BC V6T 1Z4, Canada}
\author{Debashish Mukherji}
\email[]{debashish.mukherji@ubc.ca}
\affiliation{Quantum Matter Institute, University of British Columbia, Vancouver BC V6T 1Z4, Canada}

%\date{\today}

\begin{abstract}
Heat propagation in quasi-one dimensional materials (Q1DMs) often appears paradoxical.
While an isolated Q1DM, such as a nanowire, carbon nanotube, or polymer,
can exhibit a high thermal conductivity $\kappa$, forests of the same materials show
a reduction in $\kappa$. Here, the complex structures of these assemblies have hindered the emergence of a clear 
molecular picture of this intriguing phenomenon. We combine multiscale (coarse-grained) simulation with the concepts known 
from polymer physics and thermal transport to unveil a generic (microscopic) picture of 
$\kappa$ reduction in molecular forests. We show that a delicate balance between the bond orientations, 
the persistence length of the Q1DM and the flexural vibrations govern the knock-down of $\kappa$.
\end{abstract}

\maketitle

Thermal transport properties are important for heat management in devices, 
energy generation and storage, electronic packaging, 
house-hold items, bio-materials and interfacial composites, to name a few examples \cite{cahill03jap,tomanek00prl,pipe14nm,chang17prl,zhang17prl,fuga17ps}. 
A low thermal conductivity $\kappa$ is extremely desirable for thermoelectrics \cite{balandin10nl,kodama17nm,wang17afm},
while very high $\kappa$ values are needed for heat sinking applications \cite{pipe14nm,ren11nl,cola16aami}.
The ability to tune thermal properties is thus broadly important, and nanoscale systems present a great opportunity in this regard,
because they often lead to unique and unexpected behavior \cite{mar03rmp,char07rmp}.

Understanding heat transport in quasi-one dimensional materials (Q1DMs) 
is a scientifically major problem \cite{mar03rmp,char07rmp}, and becomes even more challenging when 
dealing with arrays of Q1DMs, such as forests \cite{heattrap,cahill12jap,pan15apl,penn18nano,jakub10carbon}, bundles \cite{jakub14book}, sheets \cite{yama19apl}, 
and fibers \cite{shen10natnano,henry10prb}.
Here, the complex molecular structure of an array introduces entropic disorder and thus controls its 
physical properties. One such intriguing phenomenon is strong heat localization in carbon nanotube (CNT) forests \cite{heattrap}:
while a single CNT exhibits $\kappa_{\rm ||} > 10^3~{\rm Wm^{-1}K^{-1}}$ \cite{tomanek00prl,chang17prl,zhang17prl}, 
CNT forests show a drastic reduction in $\kappa_{\rm ||}$. Here, $\kappa_{\rm ||}$ is the thermal conductivity along the 
molecular backbone. This ``heat trap" effect was observed at a very high temperatures of $T > 10^3$ K.
However, room temperature experiments have also yielded relatively low values, such as
$\kappa_{\rm ||} \simeq 0.5-1.2~{\rm Wm^{-1}K^{-1}}$ for CNT forests \cite{jakub10carbon},
$\kappa_{\rm ||} \simeq 100~{\rm Wm^{-1}K^{-1}}$ for CNT bundles \cite{jakub14book} or
$\kappa_{\rm ||} \simeq 43~{\rm Wm^{-1}K^{-1}}$ for CNT sheets \cite{yama19apl}. We note that the observed 
influence of crowding on $\kappa_{\rm ||}$ in CNT forests and sheets may not be a system specific phenomenon.
Other examples include nanowire (NW) arrays \cite{cahill12jap,pan15apl,penn18nano}, 
polyethylene (PE) fibers \cite{shen10natnano}, crystalline-like assemblies of PE \cite{henry10prb} 
and poly-3,4-ethylenedioxythiophene (PEDOT) \cite{crn18prm}, and composite materials \cite{hashin}.

Heat transport in an isolated Q1DM has been studied extensively \cite{mar03rmp,char07rmp,dd07prl,chen08prl,ou08apl,dd09prl,lee16nl}.
A few studies have also investigated the effects of crowding on $\kappa$, where individual molecules are 
randomly orientated in a sample \cite{keb09prl,vol13apl}. In these studies, the non-bonded van der Waals (vdW) contacts 
between different Q1DMs strongly influence their thermal behavior, especially when the molecular lengths are smaller than 
the sample dimensions. This vdW-based interaction also leads to a low $\kappa$ \cite{keb09prl}.
A molecular forest, however, is inherently anisotropic, and 
a delicate balance between the bonded interactions 
and molecular entanglements dictates the behavior of $\kappa_{\rm ||}$ along the molecular orientations.
In the lateral directions, $\kappa_{\perp}$ is dominated by the weak vdW interaction.
Generally, $\kappa$ between pure bonded neighbors is about 50$-$100 times larger than that between the non-bonded neighbors \cite{shen10natnano}. 
Therefore, it is rather challenging to predict {\it a priori} how crowding can account for a large 
drop in $\kappa_{\rm ||}$ \cite{heattrap,yama19apl,pan15apl,penn18nano,henry10prb,crn18prm}. 

In this work we study the anisotropic heat flow in molecular forests
using a multiscale molecular simulation approach. 
For this purpose we: (1) devise a generic scheme to map 
the nanoscale physics onto a coarse-grained (CG) model, (2) develop a microscopic understanding of the reduced heat 
transport in molecular forests, and (3) show how a broad range of materials can be modelled within one 
unified physical concept. To achieve the above goals, we combine molecular dynamics simulations of 
a generic polymer brush model with known theoretical concepts from polymer physics \cite{doibook} and thermal transport \cite{cahill92prb,braun18am}.

We consider a Q1DM as a linear polymer chain, where the inherent flexibility is dictated by 
its persistence length $\ell_{p}$. For example, a linear molecule behaves as a rigid 
rod when the contour length $\ell_c \simeq \ell_{p}$,
while it follows a self-avoiding random walk statistics for $\ell_c >> \ell_{p}$ \cite{doibook}.
A recent experiment estimated that $\ell_p$ of an isolated single wall CNT 
is about $50-60~\mu$m for a CNT diameter of $\mathcal{D} \simeq 1.0$ nm \cite{cntlp}. 
Furthermore, $\ell_p \simeq 5~\mu$m for a NW with $\mathcal{D} \simeq 1.0$ nm \cite{sinwlp}, 
$\ell_p \simeq 0.65$ nm for PE \cite{pelp} and $\ell_p \simeq 0.5-1.5$ nm for PEDOT \cite{polh}.
Using these $\ell_p$ values, we can now analyze molecular forests. For example, 
the typical heights $\mathcal{H}$ of CNT forests or arrays of NWs range within $0.1-2$ mm \cite{heattrap,yang17sp}
and in some cases can also be 6 mm \cite{jakub10carbon}. For the bundles of PE \cite{shen10natnano} 
or PEDOT \cite{crn18prm}, $\mathcal{H} \simeq 100$ nm. 
Therefore, it is evident that $\mathcal{H} \simeq 2-200\ell_p$ in most cases. 
This observation provides an important length scale in our simulations and suggests that a long Q1DM can be
modelled as a flexible polymer chain, and hence a molecular forest as a polymer brush. Furthermore, the
(covalently) bonded monomers along a chain backbone impart almost crystalline-like structure, while the vdW 
interactions dominating in the lateral directions induce an amorphous-like two-dimensional packing. 
This is very similar to the situation in molecular forests \cite{heattrap,yama19apl,pan15apl,penn18nano,henry10prb,crn18prm}. 
It should be emphasized that, while a simple polymer model 
is certainly not appropriate to describe all the complex properties of Q1DMs, our aim is to investigate 
if a CG model can explain the anomalies in thermal behavior 
observed in experiments \cite{heattrap,jakub14book,yama19apl}. 

For this study, we employ the bead-spring polymer model \cite{kgmodel}. 
In this model, individual monomers interact with each other via a repulsive 6$-$12 Lennard-Jones potential with a cutoff 
distance $r_c = 2^{1/6}d$. $V_{\rm LJ} = 0$ for $r> r_c$. The bonded monomers in a chain 
interact with an additional finitely extensible nonlinear elastic (FENE) potential.
The results are presented in the unit of LJ energy $\varepsilon$, LJ distance $d$ and mass $m$ of individual monomers. 
This leads to a time unit of $\tau=d(m/\varepsilon)^{1/2}$. 
We consider chains of length $N_{\ell} = 500$. Note that $\ell_p$ of the fully flexible polymer model is about one bead, 
thus in our case $N_{\ell} \simeq 500\ell_p$. Here the bond length $\ell_b \simeq 0.97d$. 
Furthermore, the first monomer of every chain is tethered randomly onto a square plane 
with lateral dimensions $L_x = L_y \simeq 36.5d$ and the chains are oriented normal to the surface, in the $z$ direction. 
The surface coverage $\Gamma$ is varied up to $0.65$. 
%Here, $\Gamma$ is calculated as the ratio of the total area occupied by all tethered monomers and 
%the surface area of the plane. $\Gamma$ values chosen here are much larger than the critical grafting density 
%defined as $\Gamma^* = \left(d/2R_{\rm g}\right)^2$ with $R_{\rm g}$ being the radius of gyration of the chain. 
Periodic boundary conditions are applied in the $x~{\rm and}~y$ directions. One set of single chain
simulations have also been performed where the chain is tethered at both ends forming a
fully stretched configuration. Further simulation details are shown in the Supplementary Section S1 and Fig. S1 \cite{epaps}.

Simulations are carried out in two stages: the initial equilibration and the thermal transport calculations. 
Initial equilibration is performed under the canonical ensemble with a time step of $\triangle t = 0.01 \tau$ for
$2 \times 10^7$ MD time steps. The equations of motion are integrated using the velocity Verlet algorithm \cite{vverlet}. The system is 
thermalized via a Langevin thermostat with a damping constant $\gamma = 1\tau^{-1}$ and $T = 1\varepsilon/k_{\rm B}$,
where $k_{\rm B}$ is the Boltzmann constant. After this step, the components of $\kappa$ are calculated using the Kubo-Green
method in microcanonical ensemble \cite{greenkubo}. More details on $\kappa$ calculations are in 
the Supplementary Section S2 and Fig. S2 \cite{epaps}.

\begin{figure}[ptb]
\includegraphics[width=0.43\textwidth,angle=0]{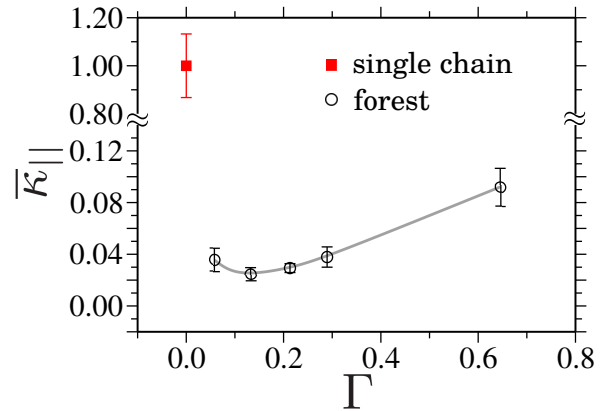}
\caption{Normalized thermal conductivity along a chain backbone ${\overline {\kappa}}_{\rm ||} = {\kappa}_{\rm ||}(\Gamma)/{\kappa}_{\rm ||}(0)$ 
as a function of surface coverage of polymers $\Gamma$. ${\kappa}_{\rm ||}(0)$ corresponds to the single chain data (i.e., $\Gamma \to 0$),
where the chain is tethered at both ends. For the simulations
under crowded environments, we have only calculated $\kappa_{\rm ||}$ of a single chain in a brush configuration, such that a 
chain experiences a cylinder-like confinement. Note that normalization volume in the Kubo-Green formula is taken as the volume of one chain,
i.e., ${v} = v_m N_{\ell}$ with $v_m$ being the volume of one monomer.
The gray line is a polynomial fit to the data that is drawn to guide the eye.
\label{fig:kappa_gamma}}
\end{figure}

\begin{figure*}[ptb]
\includegraphics[width=0.76\textwidth,angle=0]{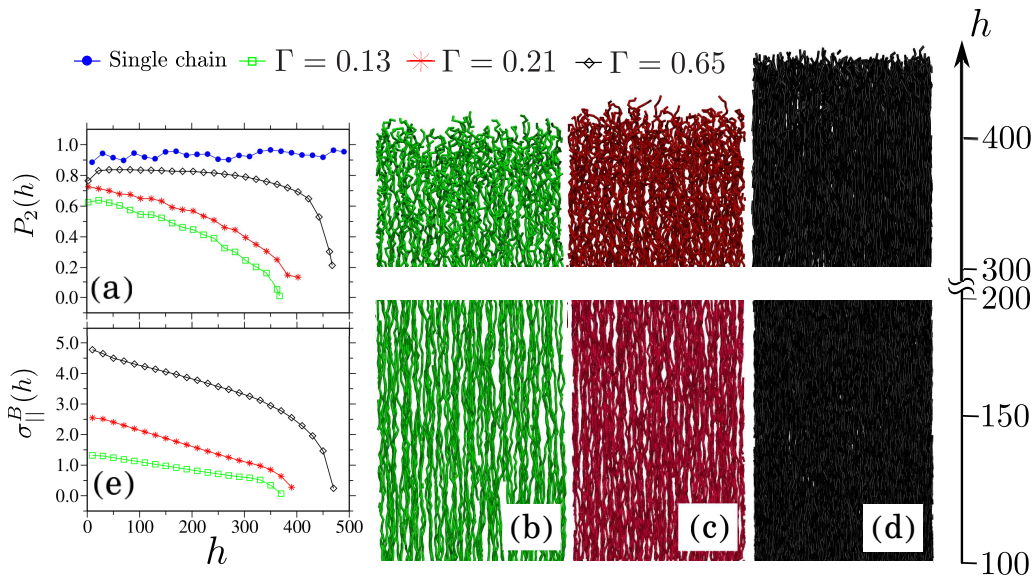}
	\caption{(a) shows the second Legendre polynomial of the bond orientation vector $P_2$ 
	as a function of the brush height $h$ along the $z$ axis. Data is shown for 
	a single chain and for three different surface coverage concentration $\Gamma$.
	(b-d) show simulation snapshots for $\Gamma = 0.13$ (green), $\Gamma = 0.21$ (red) and
        $\Gamma = 0.65$ (black). The bottom panels of the snapshots are the enlarged views of brushes
	between $100d <h< 200d$ and the top panels show the top layer for $h > 340d$.
	The arrow at the right corner points at the direction of brush height $h$.
	(e) is the same as (a) but for the stress along the chain backbone $\sigma_{\rm ||}^B$.
	The lines are drawn to guide the eye.
\label{fig:p2_snap}}
\end{figure*}

In Fig. \ref{fig:kappa_gamma} we summarize the normalized thermal conductivity along a chain backbone 
${\overline {\kappa}}_{\rm ||} = {\kappa}_{\rm ||}(\Gamma)/{\kappa}_{\rm ||}(0)$ as a function of $\Gamma$.
Here, ${\kappa}_{\rm ||}(0)$ corresponds to the single chain data (i.e., $\Gamma \to 0$).
Note that for the calculation of $\kappa_{\rm ||}$ in a brush we only consider one chain in the crowded environment.
It can be seen that, within the range $0.05 < \Gamma < 0.30$, ${{\kappa}}_{\rm ||}$ reduces by a factor of 25$-$30 in 
a brush compared to a single chain.
This sharp decrease is reminiscent of the reduced $\kappa_{\rm ||}$ in CNT forests \cite{heattrap,jakub14book} and sheets \cite{yama19apl}. 
What causes such a dramatic decrease in $\kappa_{\rm ||}$? It is particularly puzzling given that 
$\Gamma >> \Gamma^*$ in all cases (see the Supplementary Table S1 \cite{epaps}) and therefore individual chains in a brush
are expected to stretch significantly \cite{doibook}. Here, ${\kappa}_{\rm ||}$ is expected to be dominated by the bonded interactions.
In this context, a closer investigation reveals that a monomer of a chain in a crowded 
environment has two different modes of heat dissipation: 
(a) two covalently bonded neighbors and (b) $n$ non-bonded neighbors governed by the vdW interactions. 
Furthermore, the vdW interaction strength is less than $k_{\rm B}T$, 
while the bonded interactions can be typically of the order of 80$k_{\rm B}T$ (a number representative 
of a C-C covalent bond) \cite{deju,mukherji20arcmp}. The stronger bonded interaction also leads 
to about two orders of magnitude higher stiffness \cite{pe1,mukherji19prm}.
Moreover, given that $\kappa$ is directly related to the stiffness (we will come back to this point later) \cite{cahill92prb,braun18am}, 
we will now investigate how a 25$-$30 times reduction in $\kappa_{\rm ||}$ is observed in Fig. \ref{fig:kappa_gamma} and in 
experiments \cite{heattrap,jakub10carbon,jakub14book,yama19apl}. For this purpose,
we will now investigate the influence of microscopic chain conformation on $\kappa_{\rm ||}$.

We start by calculating the second Legendre polynomial $P_2$ of the bond orientation vector
using $P_2 = \left(3 \left<\cos^2(\theta)\right> - 1 \right)/2$. Here, $\theta$ is the angle of a bond 
vector with the $z$ axis and $\left<\cdot\right>$ represents the averages over all bonds and the simulation time. 
Here, $P_2 = 1.0$ when all bonds are oriented along the $z$ axis, 
$P_2 = 0.0$ when bonds are randomly oriented and $P_2 = -1/2$ when all bonds are perpendicular to the $z$ axis.
In Fig. \ref{fig:p2_snap}(a) we show the variation of $P_2$ with the forest height $h$ for three different values of $\Gamma$.   
It can be seen that $P_2 \simeq 0.92$ for a single polymer with about 5\% fluctuation.
This is expected given that a single chain is fully stretched and all bonded monomers are arranged in an almost perfect
one-dimensional crystalline-like structure along the $z$ axis.
This is also consistent with a large $\kappa_{\rm ||}$ value for a single chain (see Fig. \ref{fig:kappa_gamma}).

For $\Gamma = 0.13$ and $\Gamma = 0.21$ in Fig. \ref{fig:p2_snap}(a), it can be seen that $P_2$ decreases rather sharply with $h$, as known from 
the structure of polymer brushes \cite{binder02epje}. This is consistent with the tethering constraint that 
the chains are significantly more stretched near the
tethered points and become more randomly oriented as $h$ increases, see also simulation snapshots in Figs. \ref{fig:p2_snap}(b-c).
Furthermore, the individual chain end-to-end distances are $R_{ee}^z \simeq 370 d$ (for $\Gamma = 0.13$) and $R_{ee}^z \simeq 380 d$ (for $\Gamma = 0.21$)
(see the Supplementary Table S1 and Fig. S1 \cite{epaps}), and thus are only about 75\% of the chain contour length $\ell_c = N_{\ell}\ell_b \simeq 485 d$
for $N_{\ell} =500$. This incompatibility between $R_{ee}^z$ and $\ell_c$ indicates 
a significant chain bending (via the flexural vibrations) and thus introduces a large degree
of intra-chain entanglements, also visible from the lower panels of the simulation snapshots in Figs. \ref{fig:p2_snap}(b-c).
With increasing $h$, chain bending and entanglement become more-and-more prominent, see the upper panels 
of the simulation snapshots in Figs. \ref{fig:p2_snap}(b-c).
In this context, it is important to note that in a fully stretched chain (as in our case of the single chain), phonon-like 
wave propagation carries a heat current along the chain backbone because of the periodic arrangement of monomers. 
When entanglements appear along a chain backbone due to the 
flexural vibrations dictated by $\ell_p$ (as in the cases of $\Gamma = 0.13$ and 0.21), the longitudinal phonon propagation is 
impacted. Here, each entanglement acts as a scattering center for phonon propagation and thus reduces the phonon
mean free path. The larger the number of entanglements for a given $N_{\ell}$, the higher the resistance to heat flow,
i.e., the lower $\kappa_{\rm ||}$ (see Fig. \ref{fig:kappa_gamma}). This observation is consistent with the recent simulation study of a single 
PE chain, where it has been shown that increasing the number of kinks also decreases $\kappa$ \cite{chenkink}. 
For $\Gamma = 0.65$, bonds are significantly more oriented and also the chains are
more stretched with $R_{ee}^z \simeq 450 d$ (see Figs. \ref{fig:p2_snap}(a) and (d)), 
resulting in an approximately three fold increase of $\kappa_{\rm ||}$ in comparison to $\Gamma = 0.13$ or 0.21, see Fig. \ref{fig:kappa_gamma}

Chain bending also reduces the longitudinal chain stiffness and thus $\kappa$. 
Therefore, to achieve a better quantitative relationship between $P_2$ (or an estimate of bending), 
local stiffness, and $\kappa$, we will now look into how $P_2$ can be related to stiffness (or stress).
For this purpose, we have calculated the $h$ dependent bonded contribution to the virial stress $\sigma_{\rm ||}^B$. 
The data is shown in Fig. \ref{fig:p2_snap}(e). It can be seen that the data for $\Gamma = 0.13$ and 0.21 
not only show a rather large variation with $h$, consistent with the variation of $P_2$ in Fig. \ref{fig:p2_snap}(a), 
but that this is significantly lower than $\sigma_{\rm ||}^B \simeq 5.0\varepsilon d^{-3}$ for $\Gamma = 0.65$, see Fig. \ref{fig:p2_snap}(e).

\begin{figure}[ptb]
\includegraphics[width=0.43\textwidth,angle=0]{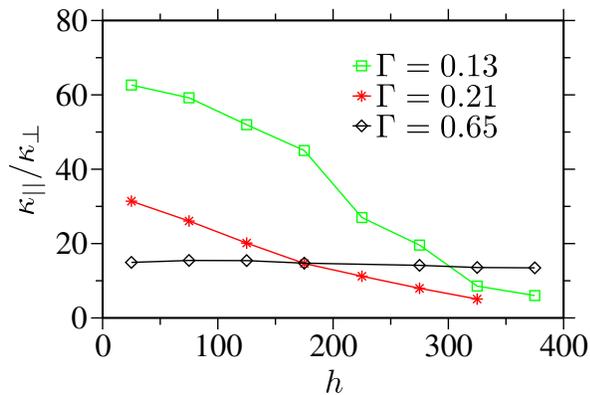}
\caption{$\kappa_{\rm ||}/\kappa_{\perp}$ as a function of the
        brush height $h$. Here, $\kappa_{\rm ||}$ and $\kappa_{\perp}$ are the 
	parallel and the perpendicular components of the thermal conductivity, respectively.
	Data are shown for three different surface coverage concentrations $\Gamma$. 
	The lines are drawn to guide the eye.
\label{fig:kratio}}
\end{figure}

Figs. \ref{fig:p2_snap}(a) and (e) also suggest that there is an inherent $h$ dependent anisotropy in the 
chain orientation, i.e., the chains are more stretched very close to $h \to 0$ due to 
tethering and become more random with increasing $h$ \cite{doibook,binder02epje}.
Consequent anisotropy is reflected in $\kappa_{\rm ||}/\kappa_{\perp}$ in Fig. \ref{fig:kratio}, 
where $\kappa_{\perp}$ is the thermal conductivity along the $x~{\rm \&}~y$ directions. 
For $\Gamma = 0.13$ we observe that heat flow is: highly anisotropic for $h < 200 d$ ($\kappa_{\rm ||}/\kappa_{\perp} \simeq 50-60$),
moderately anisotropic for $200 d < h < 350 d$ ($\kappa_{\rm ||}/\kappa_{\perp} \simeq 10-40$), and 
weakly anisotropic for $h > 350 d$ ($\kappa_{\rm ||}/\kappa_{\perp} < 10$). 
With increasing $\Gamma$, the relative anisotropy in $\kappa_{\rm ||}/\kappa_{\perp}$ decreases (see
the red and black data sets in Fig. \ref{fig:kratio}). This is predominantly because of the increased
particle number density $\rho$ that induces a faster increase in $\kappa_{\perp}$ than $\kappa_{\rm ||}$
with $\rho$ (see the Supplementary Figs. S1 \& S3 and Section S2 \cite{epaps}).
Here, considering that $\ell_p \simeq 1d$ in our model, this also gives a comparable estimate 
of the relevant length scales (in terms of $\ell_p$) that are needed to make a 
direct experimental comparison. In this context, most experiments on CNT forests deal with $\Gamma \leq 0.10$,
$1< \mathcal{D} <10$ nm (can even be several 10 nm in some cases) and also relatively small $\mathcal{H} \leq 2$ mm \cite{heattrap}. 
Therefore, the conditions typically fall within the range when $\mathcal{H}$ varies from a few $\ell_p$ to about $40\ell_p$ 
(i.e., for $\mathcal{D} \simeq 1.0$ nm and $\mathcal{H} \simeq 2$ mm) \cite{cntlp}. 
This will then lead to a rather anisotropic regime \cite{heattrap,yama19apl}.
Here, experiments on CNT forests yielded a $\kappa_{\rm ||}/\kappa_{\perp} \simeq 10-100$ \cite{heattrap,jakub10carbon},
for CNT sheets $\kappa_{\rm ||}/\kappa_{\perp} \simeq 500$ \cite{yama19apl},
and for PE fibers $\kappa_{\rm ||}/\kappa_{\perp} \simeq 1000$ \cite{shen10natnano}, 
Furthermore, our simulations show $\kappa_{\rm ||}/\kappa_{\perp} \simeq 10-60$ for $h < 200 \ell_p$ and with varying $\Gamma$, see Fig. \ref{fig:kratio}.
This further suggests that our simple CG model captures the relevant physics of the problem.

\begin{figure}[ptb]
\includegraphics[width=0.310\textwidth,angle=0]{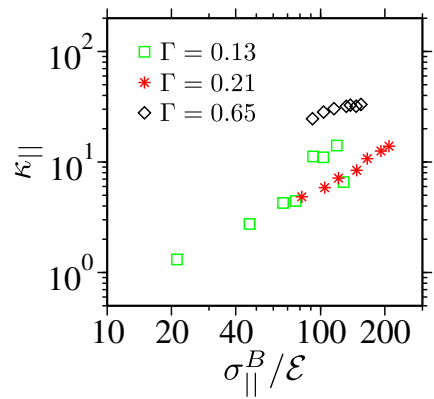}
	\caption{The parallel component of the thermal conductivity $\kappa_{\rm ||}$
	as a function of an estimated elastic modulus along the chain orientation $\sigma_{\rm||}^B/\mathcal{E}$.
	Here, $\sigma_{\rm||}^B$ and $\mathcal{E}$ are the longitudinal bonded components of the stress and 
	strain, respectively. Data are shown for three different
	surface coverage concentrations $\Gamma$.
\label{fig:stresstherm}}
\end{figure}

Lastly, Fig. \ref{fig:stresstherm} shows $\kappa_{\rm ||}$ as a function of 
an estimate of elastic modulus along the direction of the chain orientation $\sigma_{\rm||}^B/\mathcal{E}$. Here, $\mathcal{E}$
is strain. We estimate $\mathcal{E}$ from the stretching of the bond vector along the $z$ direction. %,
%which is only about 1.0\% for $\Gamma = 0.13$, 1.1\% for $\Gamma = 21$
%and 3.0\% for $\Gamma = 0.65$. These small $\mathcal{E}$ values are expected given that a bond is rather stiff \cite{kgmodel}.
It can be appreciated that the data in Fig. \ref{fig:stresstherm} is constant with an understanding that
$\kappa$ is directly related to the stiffness \cite{cahill92prb,braun18am}.

In conclusion, combining molecular dynamics simulations with known concepts from polymer physics and thermal conductivity, we
have studied the microscopic, generic behavior of anisotropic thermal conductivity in molecular forests. 
As a model system, we have used a generic coarse-grained polymer brush.
We provide a possible explanation for the reduced thermal conductivity in molecular forests, i.e., the observation that, 
while a single linear molecule can have very large thermal conductivity along the molecular backbone $\kappa_{\rm ||}$, the same molecule in a forest shows a drastic reduction in 
$\kappa_{\rm ||}$. Typical experimental systems include nanotube and nanowire forests and macromolecular fibers. 
Our analysis reveals that the reduced $\kappa_{\rm ||}$ is due to the lateral chain bending that hinders the longitudinal heat flow along the molecular backbone. 
These results point to a general principle of flexible tuning of $\kappa$ by 
changing density, molecular flexibility and forest height.
Therefore, they may pave the way towards the design of advanced functional materials with
tunable thermal properties.

%\noindent{\bf Acknowledgement:}
We thank George Sawatzky, Daniel Bruns and Manjesh Singh for useful discussions.
We further thank Celine Ruscher for a critical reading of the manuscript.
This research was undertaken thanks in part to funding from the Canada First Research Excellence Fund, Quantum Materials and Future Technologies Program.
S.P. thanks NSERC Discovery Grant program for support.
A.N. acknowledges financial support from the Natural Sciences and Engineering Research Council of Canada (Grants
No. SPG-P 478867, jointly with S.P., and No. RGPIN-2017-04608).
D.M. thanks the Canada First Research Excellence Fund (CFREF) for the financial support.
Simulations were performed at the ARC Sockeye facility of the University of British Columbia and the Compute Canada facility, which we take this opportunity to gratefully acknowledge.
Simulations in this manuscripts were performed using the LAMMPS molecular dynamics package \cite{lammps} and 
the simulation snapshots were rendered using the VMD package \cite{vmd}.

\end{document}